\documentclass{article}
\usepackage[utf8]{inputenc}
\usepackage[a4paper, total={6.5in, 9.3in}]{geometry}
\usepackage{color}
\usepackage{authblk}
\usepackage{graphicx}
\usepackage{amsmath,amssymb}
\usepackage{natbib}
\usepackage{hyperref}

\newcommand{\apj}{    {\it Astrophys. J.}}
\newcommand{\apjl}{   {\it Astrophys. J. Lett.}}

\newcommand{\pasa}{   {\it Pub. Astron. Soc. Australia}}

\chardef\us=`\_

\title{A novel algorithm for high fidelity spectro-polarimetric snapshot imaging of the low-frequency radio Sun using SKA-low precursor}

\author[1]{Devojyoti Kansabanik}
\author[1]{Divya Oberoi}
\author[2,1]{Surajit Mondal}
\affil[1]{National Centre for Radio Astrophysics, Tata Institute of Fundamental Research, S. P. Pune University Campus, Pune, India 411007}
\affil[2]{Center for Solar-Terrestrial Research, New Jersey Institute of Technology, 323 M L King Jr Boulevard, Newark, NJ 07102-1982, USA}

\begin{document}

\maketitle

\begin{abstract}
Magnetic field couples the solar interior to the solar atmosphere, known as corona. Coronal magnetic field is one of the crucial parameters which determines the coronal structures and regulates the space weather phenomena like flares, coronal mass ejections, energetic particle events and solar winds \citep{Alissandrakis2021,Vourlidas2020}. Measuring the magnetic field at middle and higher coronal heights are extremely difficult problem and to date there is no single measurement technique available to measure the higher coronal magnetic fields routinely. polarization measurements of the low-frequency radio emissions are an ideal tool to probe the coronal magnetic fields at higher coronal heights ($>1R_\odot$). Till date most of the low-frequency polarization observations of the Sun were limited to bright solar radio bursts. Here we developed a novel algorithm for performing precise polarization calibration of the solar observations done with Murchison Widefield Array, a future Square Kilometer Array (SKA) precursor. We have brought down the instrumental polarization $<1\%$. We anticipate this method will allow us to detect very low level polarised emissions from coronal thermal emissions, which will become a tool for routine measurements of global coronal magnetic field at higher coronal heights. This method can be easily adopted for future SKA and open a window of new discoveries using high fidelity spectro-polarimetric snapshot imaging of the Sun at low radio frequencies.
\end{abstract}

\section{Introduction}
%The coronal magnetic field couples the solar atmosphere to the solar interior. 
Solar magnetic fields are responsible for the bulk of the observed solar phenomenon, spanning a range of time scales from solar cycle to flares lasting milliseconds and in terms of the energetics from massive coronal mass ejections (CMEs) to nanoflares. Magnetic fields, however, are rather hard to measure and observations in visible and extreme ultra-violet (EUV) are able to provide routine estimates of magnetic fields only at low coronal heights \citep{Wiegelmann2014}.
Under favourable circumstances, radio observations have also been used to estimate coronal magnetic fields associated with active regions and/or CMEs \citep{Alissandrakis2021}. Most of the radio studies have focused on the active emissions and the global quiescent coronal magnetic fields at high coronal heights have remained beyond reach. Polarization properties of the radio emissions can provide strong constraints on the emission mechanism and serve as a direct probe of coronal magnetic fields. Despite its well appreciated importance, low-frequency polarimetric observations of the Sun are one of the least explored area of solar physics, mainly due to the technical difficulties in making these measurements. Polarimetric imaging is essential for estimating the quiescent global coronal magnetic fields by measuring the low level of polarization of the diffuse free-free emission \citep{Sastry_2009}. The variation of solar emission over small temporal, spectral, spatial and brightness temperature ($T_B$) scales imposes a requirement for high dynamic range snapshot spectroscopic imaging. The Murchison Widefield Array (MWA) \citep{Tingay2013} operating at 80 -- 300 $\mathrm{MHz}$, has a large number of antenna elements spread over a small footprint in a centrally condensed configuration and is especially well suited for high dynamic range high fidelity snapshot spectroscopic imaging. This capability has convincingly been demonstrated for spectroscopic snapshot Stokes~I imaging \citep{Mondal2019} and has enabled interesting studies \citep{Mohan2019a,Mondal2020a,Mondal2020b}. 
Here we present a general algorithm for polarimetric calibration and imaging suitable for our application and demonstrate the efficacy of its implementation on the MWA solar data. 

\section{Challenges and Limitations}
While a general prescription of polarization calibration has been available \citep{Hamaker2000}, the complexity of the problem and its compute heavy nature have restricted most commonly available implementations to make some simplifying assumptions. The challenges of low-frequency solar observations using aperture arrays are related to the large field of views (FoVs) and the nature of the instrument. A comprehensive study of a successful approach to deal with these challenges for astronomical observations is available in the literature \citep{lenc2017}. In addition to these challenges, solar observations at these frequencies have additional challenges to deal with. The large flux density of the Sun, and the wide FoVs imply that day-time calibrator observations are corrupted by solar contributions. Hence calibrators for solar observations are usually observed before sunrise or after sunset. The large time gap between the calibrator and the source observations, in addition to the difference in pointing direction, reduce the ability of calibrator observations to constrain the true state of the instrument and the ionosphere. 
This forces an increased reliance on self-calibration based methods to obtain high dynamic range solar radio images. Despite these challenges, polarization imaging have been tried using an ad-hoc approach to mitigate the instrumental polarization \citep{Patrick2019}, which we refer to as method-I hereafter. 
The ad-hoc nature of the assumptions and the use of night time calibrator observations reduce the fidelity and dynamic range of the resulting image well below the intrinsic capability of the data.
Method-I is also known to give rise to some spurious polarization for very bright solar radio bursts and canot be used for reliable detection of very low levels of circular polarization (for e.g. $\lesssim1\%$ circular polarization from the quiet Sun thermal emission \citep{Sastry_2009}).

\begin{figure*}[!t]
    \centering
    \includegraphics[trim={1cm 12cm 1.2cm 1cm},clip,scale=0.6]{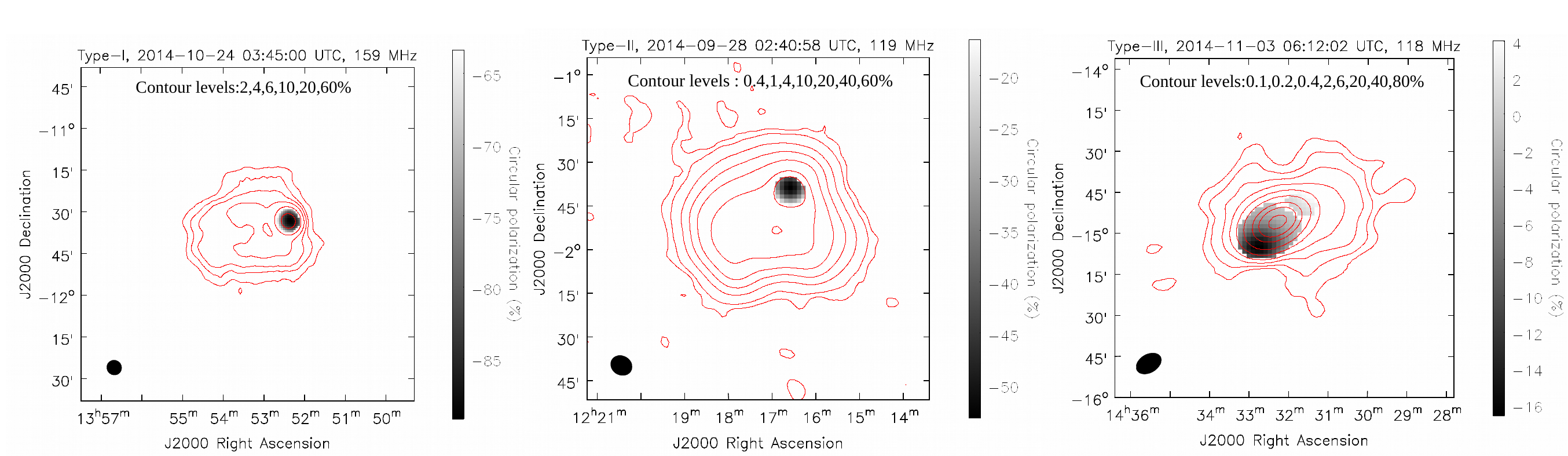}
    \caption{{\textbf{Showcase of circular polarization images for different types of solar radio bursts.}} Red contours represents the Stokes I emissions. Contour levels are at 1, 2, 4, 6, 8, 10, 20, 40, 60, 80 \% of the peak flux density. Residual Stokes V leakage in all of these images is $\lesssim2\%$ {\textbf{a.}} A type-I noise storm at 159 MHz is shown. The dynamic range of the Stokes I and Stokes V images are 788 and 518 respectively. Maximum circular polarization fraction is -89\%.  {\textbf{b.}} A type-II solar radio burst at 119 MHz is shown. The dynamic range of the Stokes I and Stokes V images are 900 and 950 respectively. Maximum circular polarization fraction is -53\%.  {\textbf{c.}} A type-III solar radio burst at 118 MHz is shown. The dynamic range of the Stokes I and Stokes V images are 12000 and 2822 respectively. Maximum circular polarization fraction is -16\%.}
    \label{fig:radio_bursts}
\end{figure*}

\subsection{Algorithm of P-AIRCARS}
\label{Overview of the Algorithm}
% brief overview and introduction to the algorithm
%This section describes our algorithm for robust polarization calibration of the MWA solar observations.
For  low  radio  frequency  solar  observations  with aperture arrays, it is not feasible to obtain calibrator observations at nearby times with the same primary beam pointings  as  used  for  target  observations, hence, this  algorithm is designed to not require any calibrator observations. We refer to this algorithm as Polarimetry using Automated Imaging Routine for Compact Arrays for the Radio Sun (P-AIRCARS).
We perform the full Jones matrix calibration using {\it CubiCal} \citep{cubical2018} which uses complex optimization and Wirtinger derivative.
P-AIRCARS builds on three pillars; i) self-calibration, ii) availability of a reliable instrumental beam model \citep{Sokowlski2017} and iii) some well-established universal properties of low-frequency solar emission. In radio interferometry, the observable is the cross-correlation between antennas $i$ and $j$, known as {\it visibility}, $V_{ij}$. 
The true $V_{ij}$s are corrupted due to instrumental and atmospheric propagation effects. It is standard practice to write the instrumental Jones matrices as a chain of independent 2$\times$2 matrices, each with its distinct physical origin and referred to as a Jones chain. The observed $V_{ij}^\prime$ is related to true $V_{ij}$ using {\it Measurement Equation} as:
\begin{equation}\label{eq:measurement_equation}
\begin{split}
 V_{ij}^\prime=&J_i\ V_{ij}\ J_j^\dagger+N_{ij},
\end{split}
\end{equation}
where $N_{ij}$ is the additive noise matrix. We estimate each of the Jones terms in the Jones chain step-by-step during the calibration process. 

\subsection{Intensity and Bandpass self-calibration}
We first estimate the time variable instrumental and ionospheric gain, $G_i(t)$, normalized over all antenna elements of the array using a single frequency channel. Making use of the compact and centrally condensed array configuration of the MWA and the very high flux density of the Sun, we estimate both antenna gains, $G_i(t)$s and the apparent source visibility, $V_{ij,Gcor}(t)$ iteratively. We start with the calibration of only the phases of $G_i(t)$s using all the baselines originating from the core antennas of the MWA. 
Once phase-only self-calibration converges, the algorithm moves to amplitude and phase self-calibration.
No assumptions are made about the polarimetric properties of $V_{ij,Gcor}(t)$. A $2\times2$ matrix calibration is performed with the only constraint that $G_i$s must be diagonal.

Since the instrumental bandpass amplitudes and phases vary across frequency, bandpass calibration is required before combining multiple spectral channels to make an image. Conventionally, instrumental bandpass is determined using standard flux density calibrator sources with known spectra. Lack of availability of suitable calibrators pushes us to rely on bandpass self-calibration. 
This, in turn, required us to find a way to deal with the degeneracy between the instrumental bandpass shape and the intrinsic spectral structure of the source (Sun). For convenience we perform the bandpass calibration over a narrow bandwidth of 1.28 $\mathrm{MHz}$ picket bands and assume that the spectral structure of the Sun to be flat over the calibration bandwidth during sufficiently quiet solar condition. The inter picket bandpass amplitudes are corrected using an independent method described by Kansabanik. D, et al. 2022 (accepted for publication in the Astrophysical Journal). Bandpass self-calibration improves the image quality and increases the dynamic range about a factor of two.

\subsection{Polarization self-calibration}
The Jones matrix corresponding to the residual phase difference between two orthogonal receptors of the reference antenna causes a leakage from Stokes U to V and vice versa. We use an image based cross-hand phase calibration with the night time observations of polarized sources relying on a method suitable for low-frequency aperture array instruments with large FoV \citep{bernardi2013}. 
Then we correct the direction dependent full Stokes ideal primary beam response of the MWA \citep{Sokowlski2017}. Method-I assumes that the leakage from Stokes I to other Stokes components remains essentially constant across the angular span of the solar disc. While this assumption is valid for some pointing directions, it is not true in general. 
We find that the percentage variation across the solar disc could be as large as 50\%. 
Hence, it is mandatory to correct for the direction dependent primary beam response for precise polarization calibration. 

Since no calibrator observation is available during solar observation, there is a degeneracy between true source polarization and residual instrumental polarization caused by the errors the ideal primary beam model, $D_i$s.
It is not possible to break this degeneracy using self-calibration. To break this degeneracy, we use a perturbative approach. 
First we perform an image based correction of the leakages based on some well known physical properties of the quiet Sun thermal emission at meter-wavelengths. These properties are: Quiet Sun thermal emission can produce a very low level of circular polarization ($\lesssim1\%$) \citep{Sastry_2009} , and no linear polarization is expected \citep{Alissandrakis2021}. We use the source model obtained after image based correction as an initial model for perturbative polarization self-calibration.

\subsubsection{Perturbative correction : Poldistortion estimation and correction}\label{sec:poldist}
Since we have already made a first order image based correction, true errors on ideal beam, $D_{i,true}$s, are expected to be close to the identity matrix. 
The small residual errors in source model can cause a deviation of the average estimated $D_{i}$ from identity.
This average deviation is referred to as poldistortion, $P_D$. 
We estimate $P_D$ by minimizing the sum of the variance of the differences of $D_{i,true}$ from identity matrix, $I$ as,
\begin{equation}
\begin{split}
 \frac{\partial S}{\partial P_D} & =\frac{\partial}{\partial P_D}\ \sum_ivar(D_{i,true}-I)\\
 P_D & =[(\sum_i D_{i}^\dagger\ D_{i})^{-1}\ \sum_i D_{i}^\dagger]^{-1}
\end{split}
\end{equation}
We then correct the visibilities using $D_{i,true}$ and obtain the final true source visibility.

\section{Results}\label{sec : result}
P-AIRCARS is the state-of-the-art polarization calibration algorithm for low-frequency solar polarimetry. 
We have tested this algorithm on several datasets from MWA under different solar conditions. 
Figure \ref{fig:radio_bursts} presents the Stokes I and V images for different solar radio bursts. Generally, P-AIRCARS obtains $\lesssim 2\%$ residual leakages for Stokes Q and Stokes U and $\lesssim1\%$ residual leakage for Stokes V. 
Such accuracy in polarization calibration is similar to what is observed for astronomical observations \citep{lenc2017} and has not reported earlier for any polarimetric solar radio studies \citep{Patrick2019}. 
We note that prior to AIRCARS \citep{Mondal2019}, the highest imaging dynamic range for solar imaging at metrewaves was a few hundred ($<$ 300) and the imaging fidelity was too poor to be able to reliably detect features of strength few percent of the peak. P-AIRCARS improves the dynamic range and image fidelity compared to AIRCARS, which leads to detection of much fainter gyrosynchrotron emissions from CMEs  (Kansabanik  et al., 2022, in preparation) compared to previous studies \citep{Mondal2020a}.  Since a robust polarization calibration is implemented in P-AIRCARS, all known instrumental effects are estimated and corrected. 
%P-AIRCARS does not have any limitation to the detection of faint polarised emissions due to any ad-hoc assumptions. 
The errors on the circular polarization fraction come primarily from the thermal rms noise of Stokes I and V images and not from uncorrected instrumental leakages or {\it ad-hoc} assumptions involved.
The rms noise for Stokes I and V images is significantly smaller than that obtained using method-I. 
Figure \ref{fig:radio_bursts} shows the Stokes I and V images from type-I, type-II, and type-III solar radio bursts. The peak circular polarization for these radio bursts is -89\%, -53\%, and -16\%,
and the peak flux densities 70, 180 and 350 SFU respectively.
Note that the peaks of the Stokes I and V emissions are not coincident for Fig. \ref{fig:radio_bursts}c.
In this image, despite the presence of the bright type-III burst emission, the high DR enables us to detect a significant part of the quiet Sun Stokes I emission, which was barely visible in the images from method-I. 

%The high DR images also allowed us to detect a significant part of the quiet Sun emission in Stokes I even the presence of the bright type-III radio bursts (Fig. \ref{fig:radio_bursts}c), which was barely visible in the images from method-I. 

\section{Conclusion}

We have developed a sophisticated and robust state-of-the-art polarization calibration algorithm tailored for the needs of low-frequency solar observations. P-AIRCARS takes advantage of the MWA design features to perform full polarimetric calibration without requiring dedicated observations of calibrator sources.
Though P-AIRCARS has been developed for polarization calibration of the solar observations, its core algorithms is general and do not impose any solar specific constraints. The perturbative approach followed here can be used for full Jones polarization self-calibration of the astronomical observations when a good initial sky model is available. 
%Polarization self-calibration was first demonstrated on simulated data in 2006 [6], but had never been used for solar observations. 
We present the first implementation of polarization self-calibration for solar observations and demonstrate its capability to achieve high dynamic range and high fidelity polarization images. 
The residual Stokes leakages for these images are on par with the usual astronomical images.
P-AIRCARS has been developed with the future SKA in mind. It can be easily adapted for unsupervised generation of high fidelity high dynamic range full Stokes images for the SKA and other similar instruments. 
We envisage that it will serve as a very useful tool for the solar and heliospheric physics community in times to come.

%We have developed a novel state-of-the-art polarization calibration algorithm for low-frequency solar observations. This algorithm is designed to work even without any calibrator observation. P-AIRCARS exploits the compact core dominated array configuration of the MWA to perform the calibration without any calibrator source. Though P-AIRCARS has been developed for polarization calibration of the solar observations, its core algorithms are independent of any solar constraints. The perturbative approach can be used for full Jones polarization self-calibration of the astronomical observations when a good enough initial sky model is available. We have first time implemented the polarization self-calibration on the solar observations and demonstrates its capability for achieving high dynamic range and high fidelity polarization images. Moreover, P-AIRCARS is developed to bear in mind about the future square kilometer array for unsupervised generation of high fidelity high dynamic range full Stokes images for SKA solar observations as well. 

\section{Acknowledgements}
This scientific work makes use of the Murchison Radio-astronomy Observatory (MRO), operated by the Commonwealth Scientific and Industrial Research Organisation (CSIRO). We acknowledge the Wajarri Yamatji people as the traditional owners of the Observatory site.  Support for the operation of the MWA is provided by the Australian Government's National Collaborative Research Infrastructure Strategy (NCRIS), under a contract to Curtin University administered by Astronomy Australia Limited. We acknowledge the Pawsey Supercomputing Centre, which is supported by the Western Australian and Australian Governments. We acknowledge support of the Department of Atomic Energy, Government of India, under the project no. 12-R\&D-TFR-5.02-0700. SM acknowledges partial support by USA NSF grant AGS-1654382 to the New Jersey Institute of Technology. We thank the developers of Python \citep{van1995python} and the various associated packages, especially Matplotlib \citep{Hunter:2007}, Astropy \citep{price2018astropy} and NumPy \citep{Harris2020}.

\bibliography{summary.bib}
\bibliographystyle{mnras}

\end{document}